\documentclass[journal]{IEEEtran}
\usepackage{amsmath}
\usepackage{amsfonts}
\usepackage{mathrsfs}
\usepackage{amssymb,color,cite,array}
\usepackage[dvips]{graphicx}

 \allowdisplaybreaks[4]

\begin{document}

\title{Multi-Antenna Covert Communications with Random Access Protocol }
\author{Weile Zhang, Nan Zhao,~\IEEEmembership{Senior Member}, Shun Zhang, and F.~Richard~Yu,~\IEEEmembership{Fellow,~IEEE}
\thanks{W. Zhang is with the School of Electronic and Information Engineering,
Xi'an Jiaotong University, Xi'an, Shaanxi, 710049, China. (email:
wlzhang@mail.xjtu.edu.cn)}
\thanks{N. Zhao is with School of Information and Communication Engineering, Dalian University of Technology, Dalian, China. (email: zhaonan@dlut.edu.cn)}
\thanks{S. Zhang is with the State Key Lab of Integrated Services Networks, Xidian University, Xi'an, Shaanxi, 710071, China (email: zhangshunsdu@gmail.com).}
\thanks{F. R. Yu is with the Department of Systems and Computer Engineering, Carleton University, Ottawa, ON, K1S 5B6, Canada (email: richard.yu@carleton.ca).}

}

 \maketitle

\vspace{-10mm}
 \begin{abstract}
In this paper, we consider the issue of covert communications with random access protocol. We consider that the legitimate user Bob has no priori knowledge about packet arrival time and thus employs data-aided frame detection based on reference sequence. The warden user Willie tries to detect this covert communication by using a radiometer. The detection performance analysis is provided for  both Bob and Willie under Rayleigh fading channel. It is demonstrated that the covert performance can benefit from the increase of the frame length or the number of receive antennas from frame detection perspective.  Numerical results are provided to verify the proposed studies.
 \end{abstract}

  \begin{keywords}
Covert communications, random access, frame detection perspective.
 \end{keywords}

\section{Introduction}

Covert communication targets at hiding wireless transmissions, which meets the ever-increasing desire of strong security and privacy.
In a typical covet communication system, a transmitter (Alice) intends to communicate with a legitimate receiver (Bob) without being detected with a warden (Willie), who is observing this communication.

 One pioneer work in~\cite{Bash13} derived a square root law for covert communication in additive white Gaussian noise (AWGN) channel, which showed that Alice can transmit no more than $\mathcal{O}(\sqrt{n})$ bits in $n$ channel users covertly and reliably to Bob.
The authors in~\cite{Lee15,He17CL} further studied the case that the warden has uncertainty about its noise power, where~\cite{Lee15} considered the worst-case approach from the warden's perspective while~\cite{He17CL} took into account the statistical distribution of noise uncertainty. The authors in~\cite{Shahzad17} considered a covert communication system under block fading and the users experience uncertainty about the channel knowledge.  Yan \textit{et. al.}~\cite{yanTIFS19} studied delay-intolerant covert communication with a finite block length and proposed to use uniformly distributed random transmit power to enhance covet communications.
Covert communication with a Poisson field of interferers has been studied in~\cite{BiaoHeTWC18}. The throughput analysis about the convert communication between Alice and Bob subject to given requirements on the covertness performance was provided based on stochastic geometry~\cite{BiaoHeTWC18}.

On the other hand, some works on covert communication in relay networks have also been reported. For example, Hu \textit{et. al.}~\cite{HuTWC18} examine the possibility and achievable performance of covert communication in one-way relay networks.
In~\cite{Jianquan19},  the channel uncertainty has been exploited to introduce confusions to degrade the performance of Willie's detection.


All of the above works have assumed that Bob exactly knows when Alice transmits. However, in a wireless system with random access protocol, even the legitimate user Bob has no exact priori knowledge about the packet arrival time, and thus the frame arrival detection is still demanded. To the best of our knowledge, the covert performance analysis with random access protocol has not been reported in the existing literatures.

In this paper, we consider the issue of covert communications with random access protocol. Bob has no priori knowledge about packet arrival time and thus employs data-aided frame detection based on reference sequence. Willie tries to detect this covert communication by using a radiometer.
The detection performance analysis is provided for  both Bob and Willie under Rayleigh fading channel.
  We show that, given required false alarm probability (FAP), when missed detection probability (MDP) of Bob keeps constant, the detection probability of Willie can be gradually decreased to its lower bound
 by increasing the frame length or the number of receive antennas at Bob.
 Numerical results are provided to verify the proposed studies.

\section{System Model}

As illustrated in Fig. 1, we consider a typical covert communication scenario where Alice attempts to transmit a packet to the legitimate receiver Bob with a low probability of being detected by warden Willie.
We assume a random access protocol is employed such that Bob has no priori knowledge about the packet arrival time and thus, the frame arrival detection is essential.
Note that in practice a data packet is usually preceded with a known sequence (the so-called reference signal or training signal). We assume that the whole packet has length $N$ and can be divided into the length-$N_{RS}$ reference  sequence and the subsequent length-($N-N_{RS}$) data sequence.

 \begin{figure}[t]
\begin{center}
\includegraphics[width=45mm]{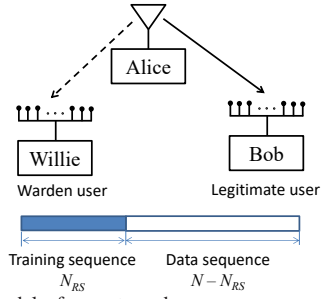}
\end{center}
\vspace{-6mm}
\caption{ System model of covert random access communications.  Both the legitimate user Bob and warden user Willie have no priori knowledge about the packet arrival time.}
\end{figure}

We consider that the legitimate user Bob has the knowledge of reference sequence, while  Willie has no knowledge about this information.
Hence, Bob can employ data-aided frame detection, while blind detection should be carried out at Willie.
The reference sequence is denoted by ${\bf s}_{RS} = [s_{RS}(0), s_{RS}(1), \cdots, s_{RS}(N_{RS}-1)]^T\in \mathbb{C}^{N_{RS}\times 1}$. The whole frame is denoted by ${\bf s} = [s(0), s(1), \cdots, s(N-1)]\in \mathbb{C}^{N\times 1}$.  There holds
$s(n) = s_{RS}(n)$ for $n=0,1,\cdots, N_{RS}-1$.
For simplicity, we assume that all transmitted symbols are of constant modulus.

Consider that Alice has one antenna while both Bob and Willie have $M$ antennas.  The propagation channels from Alice to Bob and Willie
are modeled as flat Rayleigh fading and denoted by
${\bf h}_B\in\mathbb{C}^{M\times 1}$ and ${\bf h}_W\in\mathbb{C}^{M\times 1}$, respectively. Without loss of generality, we assume each element of both ${\bf h}_B$ and ${\bf h}_W$ obey independent zero-mean unit-variance complex Gaussian distribution. Moreover, we assume Alice has no knowledge about the instantaneous channel information about the propagation channels.   We define two events, $\mathcal{H}_0$ and $\mathcal{H}_1$; Namely, $\mathcal{H}_0$: there exists no packet transmission; $\mathcal{H}_1$: Alice transmits a packet.

\section{Covert Communication Performance}

\subsection{Data-Aided Frame Detection at Bob}
In the presence of frame, the received $N_{RS}\times M$ training  signal at the correct timing point  can be expressed as
\begin{align}
\mathcal{H}_1 :  \quad  {\bf r}_B = \sqrt{P_T} {\bf s}_{RS}{\bf h}_B^T + {\bf N}_B , \end{align}
where $P_T$ denotes the transmission power at Alice and ${\bf N}_B\in\mathbb{C}^{N_{RS}\times M}$ denotes the additive white Gaussian noise (AWGN) matrix. We assume that each element of ${\bf N}_B$ obeys zero-mean complex Gaussian distribution with variance $\sigma_n^2$. Then, the average signal-to-noise ratio (SNR) can be expressed as $\rho = P_T/\sigma_n^2$.

On the other side, in the absence of frame, the received signal at Bob consists of  only noise, i.e.,
\begin{align}
\mathcal{H}_0 :  \quad  {\bf r}_B =   {\bf N}_B . \end{align}

According to the concept of generalized likelihood ratio test, we have the likelihood ratio:
\begin{align}\label{equ1}
 T_B = & \frac{ \max\limits_{{\bf h}_B,\sigma_n^2}\mathcal{L}( {\bf r}_B |\mathcal{H}_1) }{ \max\limits_{\sigma_n^2} \mathcal{L}( {\bf r}_B |\mathcal{H}_0) } \nonumber \\ = &  \frac{\max\limits_{{\bf h}_B,\sigma_n^2,P_T} \frac{1}{(\sqrt{2\pi\sigma_n^2})^{MN_{RS} } } \exp(-\frac{ \|{\bf r}_B -\sqrt{P_T} {\bf s}_{RS}{\bf h}_B^T\|_F^2 }{2\sigma_n^2})  } {   \max\limits_{ \sigma_n^2} \frac{1}{(\sqrt{2\pi\sigma_n^2})^{MN_{RS} } } \exp(-\frac{ \|{\bf r}_B  \|_F^2 }{2\sigma_n^2})   }
\end{align}
where $\mathcal{L}( {\bf r}_B |\mathcal{H}_1)$ and $\mathcal{L}( {\bf r}_B |\mathcal{H}_0)$ represent the likelihood functions under $\mathcal{H}_1$ and $\mathcal{H}_0$, respectively.
After some algebraic manipulations and remove the constants, (\ref{equ1}) can be equivalently reformulated as the following test statistic:
\begin{align}
\Lambda_B = \frac{ \| {\bf s}_{RS}^H {\bf r}_B \|_F^2 }{ N_{RS}  \|{\bf r}_B\|_F^2 }.
\end{align}


Given a threshold $\eta$, Bob claims a frame detection when $\Lambda_B >\eta_B$. We have the following Lemma:

\textit{Lemma 1:} Denote $\Phi(x) = \frac{1}{\sqrt{2\pi}} \int_{-\infty}^x e^{-t^2/2}dt $ as the cumulative distribution function (CDF) of the standard normal distribution. Given the detection threshold $\eta_B$, the MDP and FAP of Bob can be respectively expressed as
\begin{align}
& \mathcal{P}_{MDP}^{(B)} = \int_0^{+\infty} \frac{z^{M-1}e^{-z}}{(M-1)!}  \Phi\left( \frac{y(z)}{\sqrt{V(z)}} \right) dz, \label{MDPBob}\\
& \mathcal{P}_{FAP}^{(B)} = 1-F(\eta_B MN_{RS},M,1), \label{FAPBob}
\end{align}
where $y(z)  = \frac{M}{\rho}(\eta_B - \frac{1}{N_{RS}}) - (1-\eta_B)z$ and $ V(z) = \frac{  2(1-\eta_B)^2  }{\rho N_{RS} } z$, and $F(x,M,1)$ represents the CDF of the gamma distribution $\Gamma(M,1)$.
\begin{proof}
See Appendix A.
\end{proof}

\subsection{Blind Frame Detection at Willie}

The received $N\times M$ signal matrix corresponding to the whole frame at Willie can be expressed as
\begin{align}
\mathcal{H}_1 : \quad {\bf r}_W = \sqrt{P_T} {\bf s}{\bf h}_W^T + {\bf N}_W,
\end{align}
where ${\bf N}_W\in\mathbb{C}^{N\times M}$ denotes the AWGN matrix. We assume each element of ${\bf N}_W$ also obeys zero-mean complex Gaussian distribution with variance $\sigma_n^2$. Moreover, we assume Willie has perfect knowledge about the magnitude of $\sigma_n^2$.  Under $\mathcal{H}_1$, each element of ${\bf r}_W$ obeys zero-mean complex Gaussian distribution with variance $\sigma_w^2 = P_T + \sigma_n^2$.
In the absence of frame, the received signal at  Willie can be expressed as
\begin{align}
\mathcal{H}_0 : \quad {\bf r}_W =  {\bf N}_W,
\end{align}
each element of which obeys zero-mean complex Gaussian distribution with variance $\sigma_n^2$.
It is evident that the average energy of the received samples is a sufficient test statistic for blind frame detection at Willie.
Given a detection threshold $\eta_W$,  Willie considers a frame arrival when $\frac{\textrm{Tr}( {\bf r}^H{\bf r} )  }{MN} > (1+\eta_W)\sigma_n^2$.

\textit{Lemma 2:}  Denote $Q(x) = \frac{1}{\sqrt{2\pi}} \int_{x}^\infty e^{-t^2/2}dt $ as the tail distribution of the standard normal distribution. Given a threshold $\eta_W$, with a sufficiently large $M$, the detection probability and FAP of Willie can be respectively expressed as
\begin{align}
& \mathcal{P}_{DP}^{(W)} =  Q\left(\sqrt{MN} \eta_W -    \sqrt{MN}\rho  \right), \label{PDPW} \\
& \mathcal{P}_{FAP}^{(W)} = Q(\sqrt{MN}\eta_W). \label{FAPW}
\end{align}
\begin{proof}
See Appendix B.
\end{proof}

\subsection{Covert Communication Performance}

It is evident that the MDP performance depends on the SNR condition at the receiver and thus, in practice the detection threshold is usually determined based on a given required FAP. Hence, in the following, we consider that both the detection thresholds at Bob and Willie are determined according to a given required FAP $P_{FA}$.
Specifically, according to (\ref{FAPBob}), the detection threshold at Bob can be expressed as
\begin{align}\label{eta}
\eta_B = \frac{F^{-1}(1\!-\!P_{FA},M,1)}{MN_{RS}}.
\end{align}

By substituting (\ref{eta}), we can express $y(z)$ and $V(z)$ in (\ref{MDPBob}) as
 $y(z)   =   \big( F^{-1}(1\!-\!P_{FA},M,1) - M \big) \frac{1}{\rho  N_{RS}} - z$ and $V(z) = \frac{  2z }{\rho N_{RS} }$.
Then, we can rewrite (\ref{MDPBob}) into
\begin{align}
\mathcal{P}_{MDP}^{(B)} = \int_0^{+\infty} \frac{z^{M-1}e^{-z}}{(M-1)!}  \Phi\left( \frac{y(z)}{\sqrt{V(z)}} \right) dz = f_M\left(\frac{1}{\rho N_{RS}} \right)
\end{align}
where the function $f_M(x)$ parameterized by $M$ is defined as
\begin{align}
f_M(x) =  \int_0^{+\infty} \frac{z^{M-1}e^{-z}}{(M-1)!}  \Phi\left( -\frac{z - a x}{\sqrt{2xz}} \right) dz
\end{align}
where $a = F^{-1}(1\!-\!P_{FA},M,1) - M$. It is evident that $f_M(x)$ monotonically increases as $x$ increases.

Hence, given a required MDP $P_{MD}$, the minimal transmitted power at Alice can be expressed as
\begin{align}\label{PTequ}
P_T =  \frac{\sigma_n^2}{ f_M^{-1}(P_{MD}) N_{RS} },
\end{align}
where $f_M^{-1}(\cdot)$ denotes the inverse function of $f_M(\cdot)$.  It is evident that, under the same required level of MDP and FAP,  as the length of training sequence increases, the transmission power of Alice can be linearly reduced.

%


On the other side, the detection threshold at Willie can be given by
\begin{align}\label{etaW}
\eta_W = \frac{Q^{-1}(P_{FA})}{\sqrt{MN}},
\end{align}
 to achieve FAP of $P_{FA}$.
Moreover, let $\alpha = N_{RS}/N$ denotes the proportion of the training sequence in the whole frame. Then, according to (\ref{PDPW}) and (\ref{PTequ}), we obtain
\begin{align}
\mathcal{P}_{DP}^{(W)} = Q\left( Q^{-1}(P_{FA}) -     \frac{\sqrt{M }}{\alpha f_M^{-1}(P_{MD}) \sqrt{N} }  \right). \label{PDPWillie}
\end{align}

\begin{table}[t]
\centering \caption{Detection performance comparison given a required FAP $P_{FA}$ and MDP  $P_{MD}$ at Bob}
\vspace{-3mm}
\begin{tabular}  {|m{0.8cm}<{\centering\arraybackslash} m{0pt} |m{3cm}<{\centering\arraybackslash
}|m{3.5cm}<{\centering\arraybackslash}|} \hline
 Detector & \rule{0pt}{0.5cm} & Bob   & Willie
  \\  \hline
  Threshold & \rule{0pt}{0.5cm} &  $\eta_B = \frac{F^{-1}(1\!-\!P_{FA},M,1)}{MN_{RS}}$ &   $\eta_W = \frac{Q^{-1}(P_{FA})}{\sqrt{MN}}$ \\
  \hline
  False alarm & \rule{0pt}{0.5cm} & $\mathcal{P}_{FAP}^{(B)}  = P_{FA}$ & $\mathcal{P}_{FAP}^{(W)} = P_{FA}$ \\
  \hline
  Detection & \rule{0pt}{0.5cm} & $ \mathcal{P}_{MDP}^{(B)}  = P_{MD}$ &  \begin{equation*} \begin{subarray}{l}    \mathcal{P}_{DP}^{(W)} = Q\Big( Q^{-1}(P_{FA})  \\ \quad\quad\quad -     \frac{\sqrt{M }}{\alpha f_M^{-1}(P_{MD}) \sqrt{N} }  \Big)  \end{subarray} \end{equation*}  \\
  \hline
\end{tabular}
\end{table}

For comparison, we summarize the detection performance of both Bob and Willie in Table I. We can make the following important observations:
\begin{enumerate}
\item Given a fixed ratio $\alpha$, the detection probability of Willie can be decreased by increasing the frame length $N$. This theoretically indicates the covert communication performance can be improved by a longer transmission frame at Alice.  Nevertheless, it is interesting to see that the detection probability of Willie will be always upper bounded  by the required FAP $P_{FA}$.

\item  With a sufficiently large $N$, we can further rewrite (\ref{PDPWillie}) by Taylor's approximation as
\begin{align}
\mathcal{P}_{DP}^{(W)} \simeq P_{FA}  + \frac{\exp(-\frac{(Q^{-1}(P_{FA}))^2}{2}) \sqrt{M }}{\alpha f_M^{-1}(P_{MD}) \sqrt{2\pi N} }.
\end{align}
This indicates that with a large $N$, the detection probability of Willie will approach $P_{FA}$ approximately as $1/\sqrt{N}$ with increasing $N$.


 \item Given a fixed $N$, i.e., the frame length is fixed, increasing the ratio of training sequence length
     on one hand would reduce the the detection probability of Willie, on the other hand would sacrifice transmission efficiency. Hence, there exists the tradeoff between the covert performance and transmission efficiency when determining $\alpha$.

\end{enumerate}

As $M$ increases to infinity, there holds $\lim\limits_{M\to \infty} \frac{z^{M-1}e^{-z}}{(M-1)!} =  \delta(z-M)$. Then, we obtain
     \begin{align}
    \lim\limits_{M\to \infty} f_M(x) =  \Phi\left( -\frac{M(1 - \frac{a x}{M})}{\sqrt{2xM}} \right)
     \end{align}
Moreover, we have $\lim\limits_{M\to \infty}  F^{-1}(1\!-\!P_{FA},M,1)  = M$ and thus $\lim\limits_{M\to \infty} a = 0$. Then, there holds $ \lim\limits_{M\to \infty} f_M(x) =  \Phi\left( - \sqrt{\frac{M}{2x}} \right)$ and
\begin{align}
\lim\limits_{M\to \infty} f^{-1}_M(P_{MD}) = \frac{M}{ 2(\Phi^{-1}(P_{MD}))^2 }. \label{equ2}
\end{align}

By substituting (\ref{equ2}) into (\ref{PDPWillie}), the asymptotic version of the detection probability of Willie can be expressed as
\begin{align}
\mathcal{P}_{DP,asymp}^{(W)} =
Q\left( Q^{-1}(P_{FA}) -     \frac{ 2(\Phi^{-1}(P_{MD}))^2  }{ \alpha \sqrt{MN} }  \right). \label{DPasymp}
\end{align}
It is observed that the covert performance can also benefit from the increase of the number of antennas.



\section{Numerical Results}

In this section, we provide numerical results to verify the proposed studies.
We assume that the required FAP and MDP at Bob all equal $10^{-3}$, i.e., $P_{FA} = 10^{-3}$ and $P_{MD} = 10^{-3}$. The detection thresholds of Bob and Willie are determined according to the required FAP from (\ref{eta}) and (\ref{etaW}), respectively.
The transmit power of Alice is set as (\ref{PTequ}) to ensure the MDP performance of Bob. The ratio of training sequence is fixed as $\alpha=0.5$ in the following.

In the first example, we display the MDP and FAP curves of both Bob and Willie in Fig. 2 as the total frame length increases. We consider $M=16$ in this figure. It is evident that, with different frame lengths, the FAPs of both Bob and Willie as well as the MDP of Bob keep the expected level of $10^{-3}$.
Moreover, the analytical detection probability (DP) curve of Willie computed from (\ref{PDPWillie}) is also included for comparison. We see that,  the DP curve of Willie can be decreased as the frame length increases and closely matches the corresponding analytical one. This verifies our previous observation that the covert performance can be improved via increasing the frame length.

In Fig. 3, we evaluate the MDP and FAP performance as the receive antenna number increases. The results once again demonstrate the correctness of our analysis. As expected,  the covert performance can be also improved with a larger number of receive antennas.

\begin{figure}[t]
\begin{center}
\includegraphics[width=65mm]{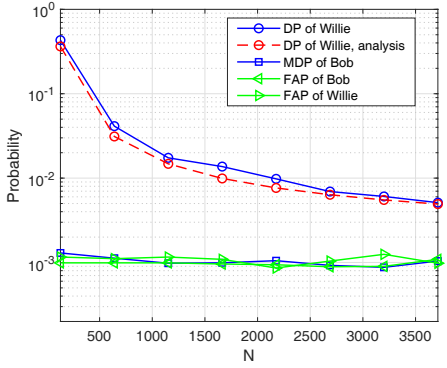}
\end{center}
\vspace{-6mm}
\caption{ MDP and FAP performance of Bob and Willie with different frame length.
}
\end{figure}

\begin{figure}[t]
\begin{center}
\includegraphics[width=65mm]{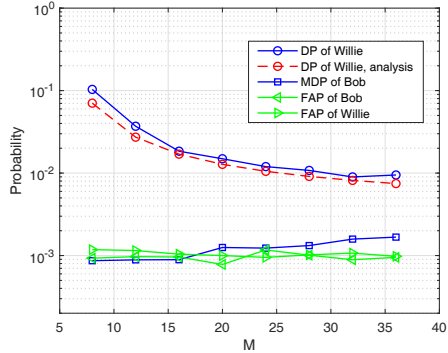}
\end{center}
\vspace{-6mm}
\caption{ MDP and FAP performance of Bob and Willie with different number of receive antennas ($M$).
}
\end{figure}

\section{Conclusions}

In this paper, we considered the issue of covert communications with random access protocol. We have derived the detection performance for both legitimate and warden users under Rayleigh fading channels. We show that the covert performance can benefit from the increase of the frame length or the number of receive antennas from frame detection perspective.

\appendices

\section{Proof of Lemma 1}

In the presence of frame, i.e., under $\mathcal{H}_1$, there hold
\begin{align}
& \| {\bf s}_{RS}^H {\bf r}_B \|_F^2 =     P_T  N_{RS}^2 \|{\bf h}_B\|^2 +  \sqrt{P_T} N_{RS} {\bf h}_B^T{\bf N}_B^H {\bf s}_{RS}  \nonumber \\ & \kern 20pt +  \sqrt{P_T} N_{RS} {\bf s}_{RS}^H{\bf N}_B{\bf h}_B^* + {\bf s}_{RS}^H {\bf N}_B{\bf N}_B^H {\bf s}_{RS}, \\
& \|{\bf r}_B\|_F^2 =  P_T  N_{RS} \|{\bf h}_B\|^2 +  \sqrt{P_T} {\bf s}_{RS}^H {\bf N}_B{\bf h}_B^*   \nonumber \\ &  \kern 20pt  +  \sqrt{P_T} {\bf h}_B^T{\bf N}_B^H {\bf s}_{RS} + \textrm{Tr}({\bf N}_B{\bf N}_B^H)
\end{align}

With a sufficient large $N_{RS}$ an $M$, we  make the approximations:
$\textrm{Tr}({\bf N}_B{\bf N}_B^H)\simeq MN_{RS}\sigma_n^2$ and ${\bf s}_B^H{\bf N}_B{\bf N}_B^H {\bf s}_B \simeq MN_{RS}\sigma_n^2$.
Given the detection threshold $\eta_B$,
the missed detection event at Bob happens when $\Lambda_B<\eta_B$, which is equivalent to the event
\begin{align}\label{appequ1}
\frac{2(1 \!-\! \eta_B)}{N_{RS}\sqrt{P_T}} \Re( {\bf s}_{RS}^H {\bf N}_B{\bf h}_B^* )
< \frac{M}{\rho} (\eta_B \!\! -\!\! \frac{1}{N_{RS}}) \!\!-\!\! (1-\eta_B)\|{\bf h}_B\|^2
\end{align}

Given one channel realization, the left hand side of (\ref{appequ1}) obeys zero-mean Gaussian distribution. We have $E[ \| {\bf s}_B^H {\bf N}_B{\bf h}_B^* \|^2 ] = N_{RS}\sigma_n^2 \|{\bf h}_B\|^2$ and thus
$E[ \Re( {\bf s}_{RS}^H {\bf N}_B{\bf h}_B^*)^2  ] = \frac{N_{RS}\sigma_n^2 \|{\bf h}_B\|^2}{2}$. Denote $z = \|{\bf h}_B\|^2$. We have $z \sim \Gamma(M,1)$.
Thus, given one channel realization $z$,  the missed detection probability can be expressed as $\Phi\big( \frac{y(z)}{\sqrt{V(z)}} \big)$, where $y(z)$ and $V(z)$ have been defined in \textit{Lemma 1}.

%
%

In the absence of frame arrival, i.e., under $\mathcal{H}_0$, the false alarm event is equivalent to
\begin{align}
\Lambda_B = \frac{ \| {\bf s}_{RS}^H {\bf N}_B \|_F^2 }{ N_{RS}  \|{\bf N}_B\|_F^2 } > \eta_B.
\end{align}
We know $\|{\bf s}_{RS}^H{\bf N}_B\|^2\sim N_{RS} \sigma_n^2 \Gamma(M,1)$. Hence,  the false alarm event happens when
$\Gamma(M,1)>MN_{RS}\eta_B$.
This completes the proof.


\section{Proof of Lemma 2}

Based on the approximation $E[\textrm{Tr}({\bf N}_W{\bf N}_W^H)] = MN\sigma_n^2$ and $Var( \textrm{Tr}({\bf N}_W{\bf N}_W^H) ) = MN\sigma_n^4$,  the detection event at Willie is equivalent to
\begin{align}\label{appequ2}
& \frac{1}{\sqrt{P_T}} {\bf s}^H {\bf N}_W{\bf h}_W^* + \frac{1}{\sqrt{P_T}}{\bf h}_W^T{\bf N}_W^H {\bf s} + \frac{1}{P_T} (\textrm{Tr}({\bf N}_W{\bf N}_W^H) - MN\sigma_n^4) \nonumber \\  & >  \frac{MN \eta_W }{\rho} -   N \|{\bf h}_W\|^2
\end{align}
The left hand side of (\ref{appequ2}) obeys zero-mean Gaussian distribution with variance $V(z) =  \frac{2N z}{\rho} + \frac{MN}{\rho^2} \simeq \frac{MN}{\rho^2} $ under low SNR region.
Then, detection probability of Willie can be expressed as
\begin{align}
\int_0^\infty  \frac{z^{M-1}e^{-z}}{(M-1)!}   Q\left( \sqrt{ \frac{N}{M}} (  M\eta_W  - z\rho )  \right) dz
\end{align}
where $y(z)  =  MN \eta_W  \sigma_n^2 - N z$.
     Given a sufficiently large $M$, we can approximate $\frac{z^{M-1}e^{-z}}{(M-1)!} \simeq \delta(z-M)$. Then, we arrive at (\ref{PDPW}).

On the other side,  in the absence of frame, according to central limit theorem, we know $\frac{\textrm{Tr}( {\bf r}_W^H{\bf r}_W )  }{MN}$ obey Gaussian distribution with mean $\sigma_n^2$ and variance $\frac{\sigma_n^4}{MN}$.
The false alarm probability at Willie can be expressed as (\ref{FAPW}).
This completes the proof.


\end{document}